\documentclass[12pt]{article}
\usepackage{epsfig}
\usepackage{latexsym}
\def\lsim{\mathrel{\rlap{\lower4pt\hbox{\hskip1pt$\sim$}}
    \raise1pt\hbox{$<$}}}         
\def\gsim{\mathrel{\rlap{\lower4pt\hbox{\hskip1pt$\sim$}}
    \raise1pt\hbox{$>$}}}         
\title{Constraints on the Variation of the Fine Structure Constant from
Big Bang Nucleosynthesis}
\author {L. Bergstr\"om, S. Iguri and H. Rubinstein\\
Department of Physics, Stockholm University\\Box 6730,
S-113 85 Stockholm, Sweden\thanks{E-mail addresses:
lbe$@$physto.se,
sergio$@$physto.se,
rub$@$physto.se.}}
\begin{document}
\date{February, 1999}
\maketitle
\begin{abstract}

We put bounds on the variation of the value of the
fine structure constant $\alpha$, at the time of Big Bang
nucleosynthesis. We
study carefully all light elements up to $^7$Li. We correct a previous
upper
 limit on
$|\Delta \alpha / \alpha|$ estimated from $^4$He primordial abundance
 and we find interesting new potential limits (depending on
the value of the baryon-to-photon ratio) from $^7$Li, whose
production is governed to a large extent by Coulomb barriers. The
presently unclear observational situation concerning the
primordial abundances preclude a better limit than $|\Delta
\alpha/\alpha| \lsim 2\cdot 10^{-2}$, two orders of magnitude less
restrictive than previous bounds. In fact, each of the (mutually
exclusive) scenarios of standard Big Bang nucleosynthesis proposed, one
based on a high value of the measured deuterium primordial abundance and one
based on a low value, may describe some aspects of data better
if a change in $\alpha$ of this magnitude is assumed.
\end{abstract}

\section{Introduction}

Physicists 
have long speculated (at least since the time of P. A. M. Dirac \cite{dirac}) about possible variations of the fundamental
physical constants. The fine structure constant, $\alpha=e^2/4 \pi$, is
especially interesting to test, being dimensionless and accurately known
experimentally.\footnote{The presently established value of $\alpha$ at
zero momentum
transfer is given by

$\alpha^{-1}=137.035989(61)$~\cite{alphanow}.}

Several attempts to constraint the time variation of $\alpha$
 have been made in the
last years~\cite{4He, lab, quasar, reactor}. The methods involved
in these computations are quite different and the results are
complementary since these calculations limit the variations of
$\alpha$ at different cosmological times. In the absence of a
particular model for the time dependence of $\alpha$, there is no
compelling reason to fit it in any particular way. Thus, it is
important to find limits on the variations of $\alpha$ at
different epochs. In particular, there may be a strong dependence
on the cosmological redshift parameter $z$ which could make most
direct measurement methods insensitive, whereas an indirect method
like the one we are investigating -- primordial nucleosynthesis --
puts us nearer the epoch of the unknown physics of the Big Bang itself
where certainly the Standard Model of particle physics is inapplicable.

Despite the lack of explicit models for the time variation of
$\alpha$, it should be kept in mind that there exist arguments,
e.g., from fundamental theories with extra dimensions~\cite{marciano,
vene, polyakov} which support the general idea of
non-constancy of the low-energy parameters. However, models in
which the coupling is governed by some condensate that varies in
time are not yet fully understood.

In some (e.g.,~\cite{vene}) the
dynamics has a ``pre Big Bang'' phase. After inflation, the universe
is such that the dilaton is already massive and the couplings are
set at the bottom of the potential without much further variation.

In another class of models~\cite{polyakov} the dilaton remain
massless but it decouples from matter. However, T. Damour and
A. M. Polyakov predict possible effects of dilaton-induced changes in
$\alpha$  and other quantities relevant to primordial nucleosynthesis to
be
of the order $10^{-8}$, though in this model other dilaton effects
are more important. In view of the primitive state of theory for
an eventual time variation of fundamental couplings, we use a
phenomenological approach and assume a different value of $\alpha$
than the present one
at the time of nucleosynthesis and investigate what the observable
consequences would be.

Direct measurements in the laboratory have given a limit on the
variation $|\Delta\alpha/\alpha|\lsim 10^{-14}$ over a period of
$140$ days~\cite{lab}. Astrophysical observations of spectra of
high red-shift quasar absorption lines have given limits of
$|\Delta\alpha/\alpha|\lsim 10^{-6}-10^{-4}$ for $z$ ranging from
$0.25$ to $3$~\cite{quasar}. The geological limit from the Oklo
natural nuclear reactor is about
$|\Delta\alpha/\alpha|\lsim 10^{-7}$ over a period of $1.8$
billion years~\cite{reactor}. Recently, it was argued that from future
observations of fluctuations in  the cosmic microwave background
radiation, variations of $\alpha$ could be bound by
$|\Delta\alpha/\alpha|\lsim 10^{-3}-10^{-2}$ for $z
\sim 10^3$~\cite{cmbr}. Finally, assuming a particular model for the
$\alpha$-dependence of the neutron-proton mass difference, a limit
can be extracted from the $^4$He primordial abundance for $z \sim
10^9-10^{10}$~\cite{4He}.

This work deals with contraints from nucleosynthesis, which have the
advantage of probing the earliest cosmological epoch where data exist and where the basic
physical processes are known from laboratory experiments work.
Our analysis corrects that of~\cite{4He} and extends it since we have
not only considered the $^4$He abundance but also the abundances
of other light nuclei.

Besides consituting a more thorough analysis of the
nucleosynthesis bounds, our treatment has the advantage of bypassing the difficult
theoretical problem of the dependence of the neutron-proton mass
difference on $\alpha$, relying on the fact that the other
abundances are only weakly sensitive to this mass difference. The
abundance of the other light elements as a function of $\alpha$
can be extracted with smaller ambiguity, and  the dependence is in
addition very steep, in particular for $^7$Li (at least in certain
ranges of the baryon-to-photon ratio, $\eta$). Of course, at the
present time these advantages are to a large extent balanced by
the disadvantage of observationally less well known abundances, something
that may hopefully improve in the future.

In the next Section we review the most important aspects of the
Standard Big Bang Nucleosynthesis (SBBN) model. Then, in Section 3
we explain how $\alpha$ enters into the SBBN scenario. In
Section 4, we study the $\alpha$-dependence of the relevant
nuclear reactions involved in primordial nucleosynthesis. Finally,
in Section 5 we estimate the corresponding limit placed on
$|\Delta\alpha/\alpha|$ at $z \sim 10^9-10^{10}$. We find that,
given the present observational uncertainties of light-element
abundances, it is not possible to put a more stringent bound than
\begin{equation}
\label{eq:alphabound}
{|\Delta\alpha|\over \alpha}\lsim 2\cdot 10^{-2},
\end{equation}
two  orders of magnitude less restrictive than the one claimed in
\cite{4He}.

\section{The fine structure constant in the SBBN scenario}

In order to give a complete account of the effects of a variation of the
fine structure constant on the primordial abundances, we will briefly
review the SBBN model. We will focus our attention on those aspects
concerning the evolution of the abundances of light nuclei.

The production of light elements according to the SBBN model is commonly
divided into three stages
\cite{turner, SKM}.
\begin{itemize}
\item First Stage: \emph{Statistical equilibrium} ($T \gg 1$~MeV; $t \ll
1$~sec)
\end{itemize}

During this first stage (as in the other two), the universe is
radiation-dominated. The relativistic degrees of freedom are
photons, electrons, positrons and the three light neutrino
species. The weak interactions that interconvert
 neutrons and protons, i.e.,
\begin{equation}
\label{eq:n>pev}
n\rightleftharpoons p+e^-+\overline{\nu}_e
\end{equation}
\begin{equation}
\label{eq:nv>pe}
n+\nu_e\rightleftharpoons p+e^-
\end{equation}
and
\begin{equation}
\label{eq:ne>pv}
n+e^+\rightleftharpoons p+\overline{\nu}_e
\end{equation}
are rapid enough to keep them in statistical equilibrium. The
neutron-to-proton ratio is, then, given by its equilibrium value (we
use units such that the speed of light, Planck's constant and
Boltzmann's constant are all set to unity) :
\begin{equation}
\label{eq:n/p|eq}
\left(\frac{Y_n}{Y_p}\right)_{eq}=e^{-\Delta m/T}
\end{equation}
where $\Delta m=m_n-m_p$ is the neutron-proton mass difference.

Moreover, at this temperature, not only are the rates for the weak
reactions~(\ref{eq:n>pev})--(\ref{eq:ne>pv}) faster than the universe
expansion rate, but so are the
nuclear reaction rates responsible for the production of light elements.
Light nuclei are then both in kinetic and chemical equilibrium or
nuclear statistical equilibrium (NSE), their corresponding abundances
being given by~\cite{turner}
\begin{eqnarray}
\label{eq:nseabun}
Y_A=&g_A \left[ \zeta(3)^{A-1} \pi^{(1-A)/2} 2^{(3A-5)/2} \right]
A^{\frac{5}{2}} \left( \frac{T}{m_N} \right) ^{3(A-1)/2}& \times
\nonumber\\ & \eta ^{A-1} Y_p^Z Y_n^{A-Z} \exp (B_A/T)
\end{eqnarray}
where $\eta$ is the present baryon-to-photon ratio, $m_N$ is the nucleon
mass, $B_A$ is the
binding energy, $g_A$ counts the number of degrees of freedom of the
nuclear species $(A,Z)$ and $\zeta$ is the Riemann zeta
function.\footnote{The numerical value
of the Riemann zeta function is $\zeta(3)\sim 1.202$.}

Due to the fact that the number density of photons is so large
relative to that of baryons ($\eta \sim 10^{-10}$), the
abundances of composite elements are completely negligible and
their synthesis does not truly start at this epoch.

\begin{itemize}
\item Second Stage: \emph{Neutron-proton ``freeze-out''} ($T \sim
0.8$~MeV; $t \sim 2$~sec)
\end{itemize}

After neutrino decoupling, at about the time
electron-positron pairs
annihilate, the second stage of primordial nucleosynthesis takes place.
The weak reactions~(\ref{eq:n>pev})--(\ref{eq:ne>pv}) become slower than the
expansion rate of the
universe and the neutron-to-proton ratio is no longer able to track its
equilibrium value: it ``freezes-out''.

After this freeze-out, the neutron-to-proton ratio can be approximated
by
\begin{equation}
\label{eq:n/p|fo}
\left(\frac{Y_n}{Y_p}\right)_{f}=e^{-\Delta m/T_{f}}
\end{equation}
where $T_{f}$ is the freeze-out temperature, which is determined by
setting the equality between the expansion and the weak rates. Note
that, since the weak rates depend on $\Delta m$, also $T_f$ depends implicitly on it.

But while reactions~(\ref{eq:n>pev})--(\ref{eq:ne>pv}) are too slow to track
the rate of expansion,
nuclear reactions are still
fast enough to keep light elements in NSE and, thus, their abundances are
still very small. It is not until the third and last stage that nuclear
production effectively begins.

\begin{itemize}
\item Third Stage: \emph{light-element synthesis} ($0.6$~MeV $\gsim T
\gsim 0.05$~MeV; ~$3$~sec~$\lsim t \lsim 6$~min)
\end{itemize}

Shortly before this time, electron-positron pairs have finally
annihilated, transferring their entropy to the photons and
increasing their temperature relative to that of the neutrinos by
a factor $(11/4)^{1/3}$. So, at this last stage the only
relativistic species are the neutrinos and the photons.

The neutron-to-proton ratio does not remain really constant after the
freeze-out but it continues to decrease during this stage from its
freeze-out value because of neutron decay and the effect of the strong
neutron-proton reactions. Actually, it will only be  at $T \sim
0.05$~MeV, when practically all available neutrons are bound into
nuclei, that the neutron-to-proton ratio  becomes constant.

But the most important feature of this period is that, while in the
previous stage only weak processes were relevant, at this epoch strong and
electromagnetic interactions get important and nucleosynthesis
 begins.

The evolution of light-element abundances is dominated by the
competition between the nuclear reaction rates and the  expansion
rate. Although the densities of the light ``fuels'' for the
reactions involved in the process are now significant, they
eventually are not big enough to keep up with the demand for the
NSE of heavier elements. Moreover, Coulomb-barrier suppression
becomes gradually  more important. Both effects result in the
freeze-out of nuclear reactions and the consequent series of
departures of light nuclei from  their NSE states.

Let us describe the process in detail, following~\cite{SKM}. The most
abundant nucleus, $^4$He, is to an excellent approximation
produced only through the mass-3 nuclei, $^3$He and tritium ($t$),
and it is only through these nuclei and the reactions
$^3$He$(n,\gamma)^4$He and $t(p,\gamma)^4$He that it is allowed to
be in NSE. At temperatures greater than $0.6$~MeV, the $^3$He and
$t$ abundances are sufficiently large to allow $^4$He to track its
equilibrium value. But at $T \sim 0.6$~MeV, the NSE curves of
the mass-3 nuclei cross with that of $^4$He, with the $^4$He
abundance rising faster than those for $^3$He and $t$. At about
this temperature, also, the reactions that maintain $^4$He in
equilibrium become too slow, $^4$He is forced to leave its NSE
curve and  follows instead the corresponding $^3$He and $t$ NSE
tracks.

This goes on until $T \sim 0.2$~MeV. At this temperature $^3$He
and $t$ also encounter a bottleneck, a ``minor'' deuterium ($d$)
bottleneck: reactions $d(n,\gamma)t$ and $d(p,\gamma)t$, which
keep $d$ in NSE with the mass-3 nuclei slow down and now $^4$He,
$^3$He and $t$ follow the $d$ NSE curve.

At $T \sim 0.08$~MeV,
the $^3$He$(n,p)t$ reaction freezes-out and the mass-3 nuclei lose
their NSE too.

Finally, all the elements encounter the ``major" deuterium
bottleneck at $T \sim 0.06$~MeV and after that the nuclear
species evolve in quasi-statistical equilibrium.

Some $^7$Li and some $^7$Be are synthesized, but due to the
existence of  energy gaps among stable nuclei at mass numbers
$A=5$ and $A=8$ and significant Coulomb-barrier suppression  at
this time, the production of nuclei beyond $A=8$ is inhibited.

\section{The $\alpha$-dependence of element abundances}

We are now able to discuss  which are the most important
$\alpha$-dependent magnitudes that affect primordial abundances.

During the first two stages of SBBN, the only $\alpha$-dependent
parameters are the weak reaction rates~(\ref{eq:n>pev})--(\ref{eq:ne>pv}),
which in turn determine the the freeze-out temperature $T_F$ (see
Eq.\,~(\ref{eq:n/p|eq})), the neutron-proton mass difference $\triangle m$
(see Eq.\,~(\ref{eq:n/p|eq})) and the
binding energies $B_A$ (see Eq.\,~(\ref{eq:nseabun})).

 Since the abundances of composite nuclei during these stages are by
themselves
negligible, the effects of a small variation in their binding energies are
 negligible too.

But this is not the case with the weak
reactions~(\ref{eq:n>pev})--(\ref{eq:ne>pv}) and with the neutron-proton mass
difference. Since almost all available neutrons after freeze-out are
finally bound into $^4$He, in a good approximation its abundance is given
by
\begin{equation}
\label{eq:4Heabun}
Y_4 \sim 2 \frac{\left(\frac{Y_n}{Y_p}\right)_{f}}{1 +
\left(\frac{Y_n}{Y_p}\right)_{f}}
\end{equation}

From this equation and Eq.\,~(\ref{eq:n/p|fo}) it follows that the main
parameters that fix the final value of $Y_4$ are precisely $\Delta m$ and
$T_{f}$.
Radiative and Coulomb corrected
expressions for the weak reactions~(\ref{eq:n>pev})--(\ref{eq:ne>pv}) have
been calculated in
\cite{radiative} and it was shown that their corrections to
primordial abundances are very small.\footnote{The
most important
correction is that for $^4$He, of about $1\%$.} The effect on $T_f$ 
of a variation
in $\alpha$ is then given only through the variation of $\Delta
m$ (see \cite{radiative, code}). That means that $Y_4$ depends just on a
single $\alpha$-dependent parameter, i.e., $\Delta m$.

This fact makes the constraint on $\alpha$ estimated from the
$^4$He primordial abundance very model-dependent: a particular
model for the $\alpha$-dependence of $\Delta m$ is needed to
estimate the change in $Y_4$ due to a change in $\alpha$. One may also
note that changes in the other gauge coupling constants, such as 
the strong coupling $\alpha_s$, may induce changes that are even more important,
but more difficult to compute than those of a changing $\alpha$
\cite{dixit,olivecamp}.

As an example of the $\alpha$-dependence of $\Delta m$, 
we have used the treatment of J. Gasser and H. Leutwyler
\cite{gasser}, from which we deduce
\begin{equation}
\label{eq:deltam}
 \Delta m \simeq 2.05 - 0.76 \cdot \left( 1 + \Delta \alpha/ \alpha \right)
{\rm~MeV}
\end{equation}

This method is a phenomenological way to evolve the mass of the
nucleon as determined by the change of quark masses due to their
strong and electromagnetic energy and the electromagnetic binding.
It is a qualitatively reasonable method but has no fundamental QCD
backing.

Other abundances are not so sensitive to $\Delta m$ as $Y_4$:
because of its large binding energy, $^4$He acts as a ``sink''
during primordial nucleosynthesis, its abundance is less sensitive
to changes in the nuclear reaction rates than the other
abundances, and more sensitive to variations in the parameters
that fix the number of neutrons relative to the number of protons,
i.e., $\Delta m$ and $T_{f}$. We note, following~\cite{dixit},
that the dimensionful weak (Fermi) coupling constant $G_F$ does
not depend on the gauge coupling constant in the standard
electroweak model.

This means that we should use the other light-element abundances
to limit $\Delta \alpha$, in addition to the $^4$He abundance,
bypassing the problem of the $\alpha$-dependence of $\Delta m$.

Our next step is, then, to analyze the role of $\alpha$ during the third
stage. The fine structure constant affects the third stage because it
enters into the expressions of the nuclear reaction rates: as we have
already mentioned, Coulomb-barrier penetration is a determining factor
during
nucleosynthesis. To implement the changes in $\alpha$ in these expressions
we first need to know exactly the way by which thermonuclear reaction
rates are obtained from experimental data. This will be discussed in the
next section.

\section{Thermonuclear reaction rates}

The rate of a nonrelativistic nuclear reaction taking place in a
nondegenerate environment is as usual given as the thermal average of the
product of the corresponding cross section $\sigma (E)$ and the relative
velocity times the number densities of the particles involved:
\begin{equation}
\label{eq:reacrate}
R_{ij}=n_in_j\langle \sigma |v| \rangle
\end{equation}

Under the assumption that these particles have isotropic Maxwell-Boltzmann
kinetic energy distributions this average can be written as~\cite{rates1}
\begin{equation}
\label{eq:sigmav}
\langle \sigma |v| \rangle=\left( \frac {8}{\mu \pi} \right)^{\frac{1}{2}}
T^{-\frac{3}{2}} \int_0^{\infty} E \sigma (E) e^{-\frac {E}{T}} dE
\end{equation}
where $\mu$ is the reduced mass.

For charged-particle induced reactions\footnote{We are not concerned
with
neutron-induced reactions since they do not involve any Coulomb barrier
penetrability factor in their cross sections and, therefore, they are not
very
sensitive to $\alpha$~\cite{neutron}.} the cross section is given by
\begin{equation}
\label{eq:chargedsigma}
\sigma (E)=\frac{S(E)}{E} e^{-2 \pi \eta(E)}
\end{equation}
where $S(E)$ is the cross section factor or the astrophysical $S$-factor
and
$\eta(E)$ is the Sommerfeld parameter:
\begin{equation}
\label{eq:eta}
\eta (E)=\sqrt{\frac{E_G}{4 \pi^2 E}} =\alpha Z_i Z_j
\sqrt{\mu\over 2E}
\end{equation}
where $E_G$ is the Gamow energy and $Z_i$, $Z_j$ the electric charges of
the
colliding nuclei.

The reason for introducing this factorization for the cross section
is the fact that, because of the exponential energy dependence of
the Coulomb barrier penetrability, charged-particle cross sections are
extremely difficult to measure at low energies. Therefore, it is necessary
to extrapolate $\sigma (E)$ from experimental data to lower energies.
Since
the
exponential factor is given by solid quantum mechanical principles, only
the
unknown nuclear physics part $S(E)$ has to be fitted, and it is
generally a slowly varying function of energy.

Let us first consider the non-resonant terms of the reactions
1--12 (see tables~\ref{tab:summary} and~\ref{tab:summary2}). Because of
its slow variation with
energy, we can expand $S(E)$ as a Taylor series:
\begin{equation}
\label{eq:Sfactor}
S(E)=S(0)+ \left. \frac{dS}{dE} \right|_{E=0} E+\frac{1}{2} \left.
\frac{d^2S}{dE^2} \right|_{E=0} E^2
\end{equation}

When inserted into Eq.\,~(\ref{eq:sigmav}) this gives
\begin{equation}
\label{eq:nrsigmav}
\langle \sigma |v| \rangle_{nr}=\left( \frac {8}{\mu \pi}
\right)^{\frac{1}{2}}
T^{-\frac{1}{2}} \sum_{i=0}^{2} \frac{T^i}{i!} \left. \frac{d^i
S}{dE^i} \right|_{E=0} N_i(\xi)
\end{equation}
where
\begin{equation}
\label{eq:Ni}
N_i(\xi)= \int_{0}^{\infty} y^i e^{-y} e^{-\xi y^{-\frac{1}{2}}} dy
\end{equation}
and where
\begin{equation}
\label{Xi}
\xi=2 \pi \alpha Z_i Z_j \sqrt{\mu\over 2T}.
\end{equation}

Introduce further $\kappa\equiv \xi^2/4$.
At low temperatures compared with the Coulomb threshold, these
``astrophysical'' integrals
are given by~\cite{rates3}
\begin{equation}
\label{N0}
N_0(\xi)=2 \sqrt{\frac{\pi}{3}} \kappa^{\frac{1}{6}}
e^{-3 \kappa^{\frac{1}{3}}} \left(
1 + \frac{5}{36} \kappa^{-\frac{1}{3}} \right)
\end{equation}
\begin{equation}
\label{eq:N1}
N_1(\xi)=2 \sqrt{\frac{\pi}{3}} \kappa^{\frac{1}{2}}
e^{-3 \kappa^{\frac{1}{3}}} \left(
1 + \frac{35}{36} \kappa^{-\frac{1}{3}} \right)
\end{equation}
and, finally
\begin{equation}
\label{N2}
N_2(\xi)=2 \sqrt{\frac{\pi}{3}} \kappa^{\frac{5}{6}}
e^{-3 \kappa^{\frac{1}{3}}} \left(
1 + \frac{89}{36} \kappa^{-\frac{1}{3}} \right)
\end{equation}

From these expressions it is straightforward to obtain the
$\alpha$-dependence of the non-resonant contributions to the reaction rates we are treating. We have not
considered the
$\alpha$-dependence of the reduced mass $\mu$ and of $S(E)$ and its
derivatives about the zero energy since this dependence obeys a
polinomial
law on $\alpha$~\cite{rates1} and it can be safely neglected.

For reactions 3, 5, 7 and 8 the non-resonant terms are predicted
to have a combination of polynomial and decreasing exponential
terms~\cite{rates1}; so, expressions like
\begin{equation}
\label{eq:Sfactorxexp}
S(E)=S(0) q(E) e^{- \beta E}
\end{equation}
where $q(E)$ is, again, a slowly varying function of the energy, must
be
included into the $S$-factor. For most of the cases we are considering, a very good approximation results when
$q(E)$ is
assumed to be a
constant, i.e., $q(E)=q(0)=1$. Then an additional term of the form
\begin{equation}
\label{eq:dsigmav}
\langle \sigma |v| \rangle_{ed}=\left( \frac {8}{\mu \pi}
\right)^{\frac{1}{2}}
T^{-\frac{3}{2}} T_{\beta} S(0) N_0(\xi_{\beta})
\end{equation}
where $T_{\beta}=T/(1+\beta T)$ and $\xi_{\beta}=2 \pi \alpha Z_i Z_j
(\mu /2)^{1/2} T_{\beta}^{-1/2}$, appears in the expressions of the corresponding reaction rates.

However, for reaction 3, $\beta$ is assumed to be of the form
\begin{equation}
\label{eq:beta}
\beta=\gamma_0 + \gamma \left( \frac {\xi}{2} \right)^{\frac{2}{3}} T
\end{equation}
and then the effective temperature $T_{\beta}$ is also an
$\alpha$-dependent parameter:

\begin{equation}
\label{eq:Tbeta}
T_{\beta}=\frac{T}{1+\gamma_0 T + \gamma \left( \frac {\xi}{2}
\right)^{\frac{2}{3}} T^2 / (1+\gamma_0 T)^{2/3}}
\end{equation}

Continuum and/or narrow and broad resonant terms appear
in the expressions for the rates of reactions 4--6, 11 and 12.
These terms are generally of the form~\cite{rates1}
\begin{equation}
\label{ressigmav}
\langle \sigma |v| \rangle_{res}= g(T) e^{-\frac{\overline{E}}{T}}
\end{equation}
where $\overline{E}$ is about the continuum threshold energy or the
resonance energy, respectively, and $g(T)$ is a function of the temperature. But since the electromagnetic
contribution to these energies is very small compared with the strong
contribution, we can safely neglect their $\alpha$-dependence.

Finally, reactions 4, 5, 11 and 12 have in addition a cut-off factor for the non-resonant terms of the
form
\begin{equation}
\label{eq:cutoff}
f_{co}=e^{- \left( \frac{T}{T_{co}} \right)^2}
\end{equation}
where $T_{co}$ is a cut-off temperature. From~\cite{rates1} it follows
that this
temperature is roughly proportional to $\alpha^{-1}$.

In tables~\ref{tab:summary} and~\ref{tab:summary2} we have explicitly
shown the relevant dependence on $\alpha$
for
all the charged-particle induced reactions having an impact on primordial abundance changes. For the relative
change in the value of 
$\alpha$, we define the quantity $\delta$:
\begin{equation}
\label{eq:delta}
\delta=\frac{\Delta \alpha}{\alpha}
\end{equation}

\section{Constraints on the variation of the fine structure constant}

Using the theoretical framework we have described we proceed to
see what is the effect of varying $\alpha$ on the relative
abundances. We do use the newest version of the SBBN code
\cite{code} to implement these variations. We keep all other
coupling constants fixed and assume no other ``exotic'' effects are
present, like strong primordial magnetic fields. We discuss these
effects below. We have performed several tests to see that the
code is working correctly. We use a neutron lifetime of
$\tau_n=886.7$~sec, which is the currently
accepted value according to the Particle Data Group~\cite{alphanow}.

\begin{figure}
\centerline{\epsfig{file=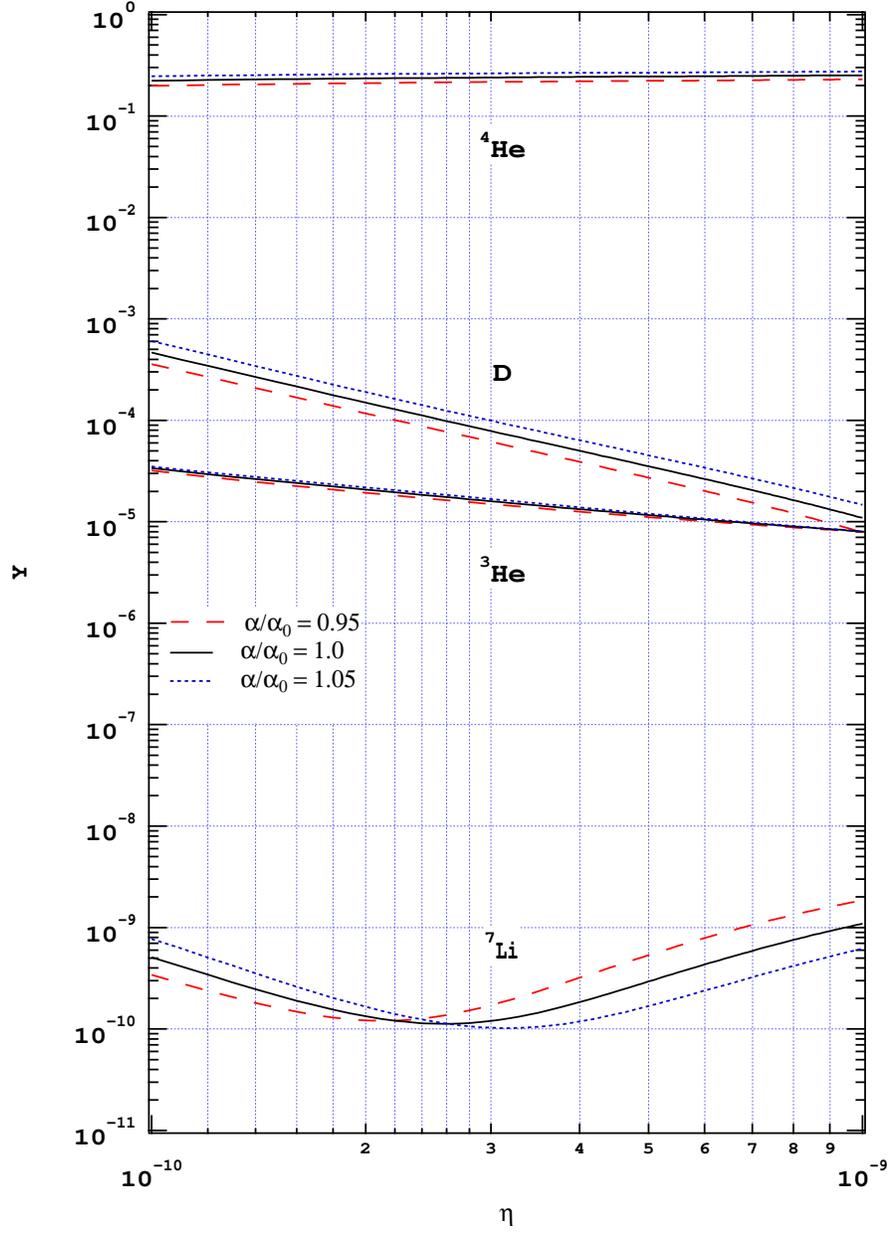,width=0.9\textwidth}}
\caption{The abundances of the light elements $D$, $^3$He, $^4$He and
$^7$Li as a function of $\eta$, the baryon-to-photon ratio. Curves
are shown for the standard value of the fine structure constant $\alpha$,
and for a variation of $\pm$ 5 \% of the standard value. The results for
the individual abundances are shown on an enlarged vertical scale in 
Figs.~\protect\ref{fig:dvseta}--\protect\ref{fig:li7vseta}.}
\label{fig:yvseta}
\end{figure}

\begin{figure}
\centerline{\epsfig{file=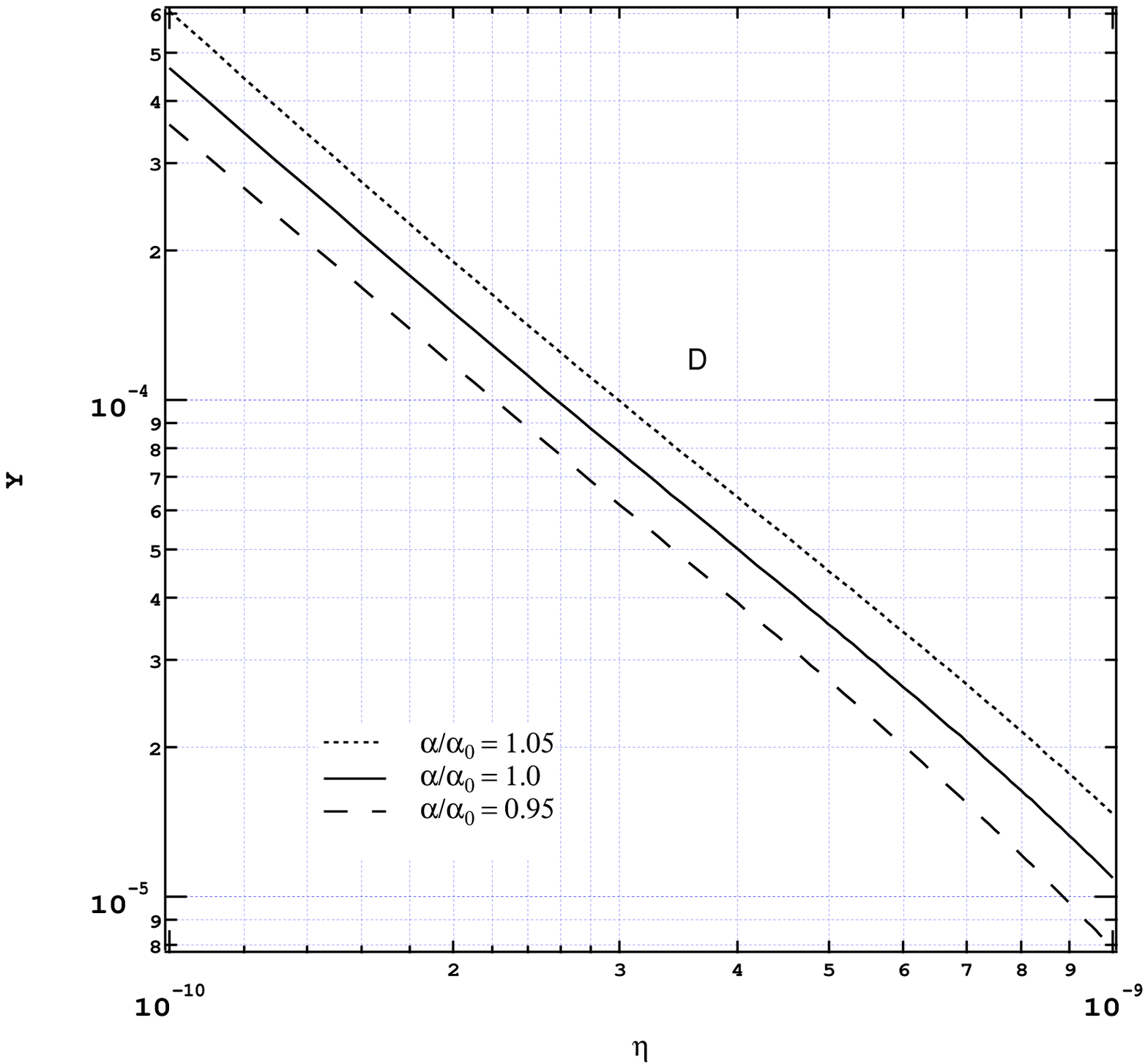,width=0.5\textwidth}\epsfig{file=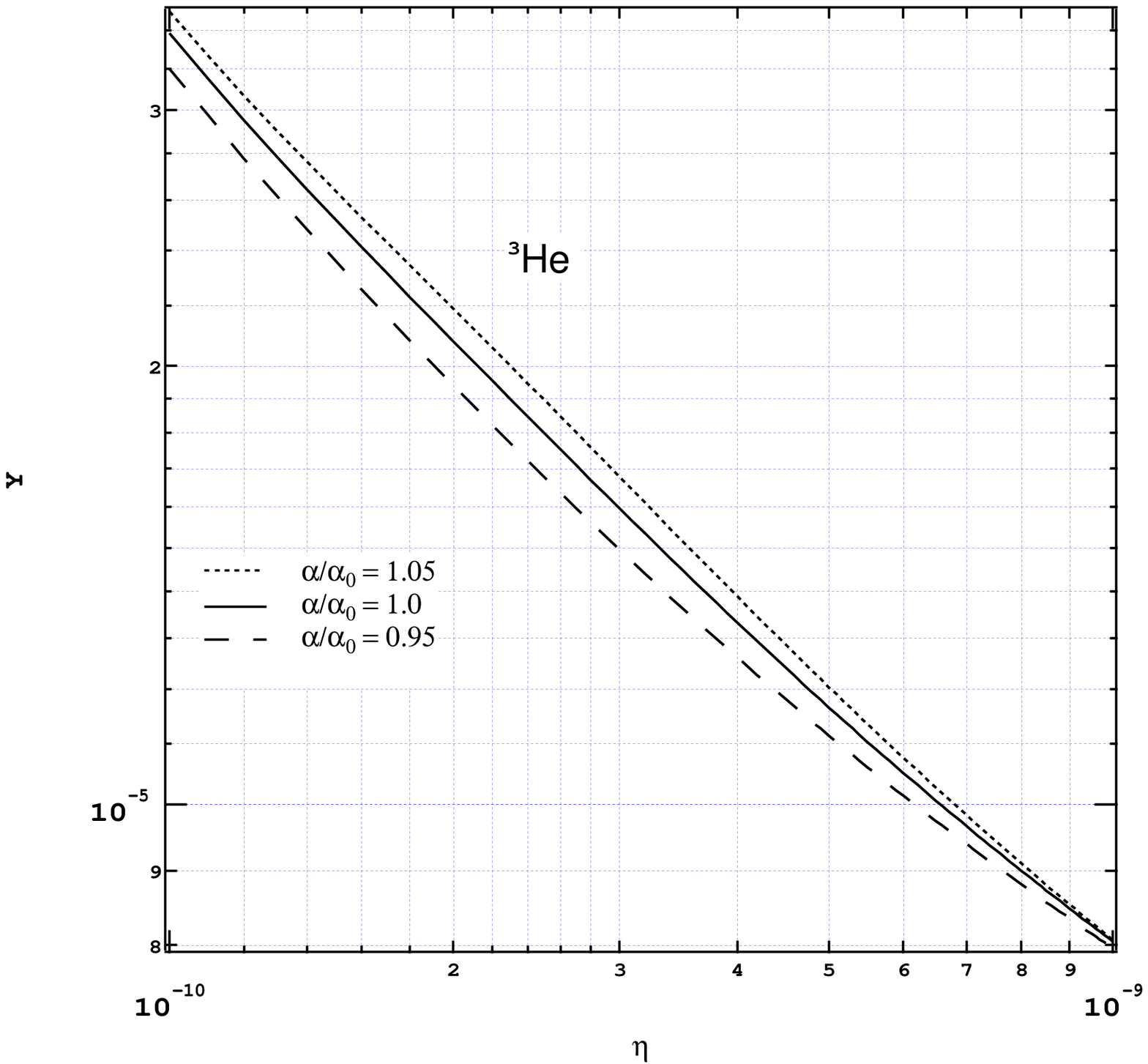,width=0.5\textwidth}}
\caption{The abundance of (a) deuterium
 and (b)  $^3$He
as a function of $\eta$, the baryon-to-photon ratio. Curves
are shown for the standard value of the fine structure constant $\alpha$,
and for a variation of $\pm$ 5 \% of the standard value.
}
\label{fig:dvseta}
\end{figure}

\begin{figure}
\centerline{\epsfig{file=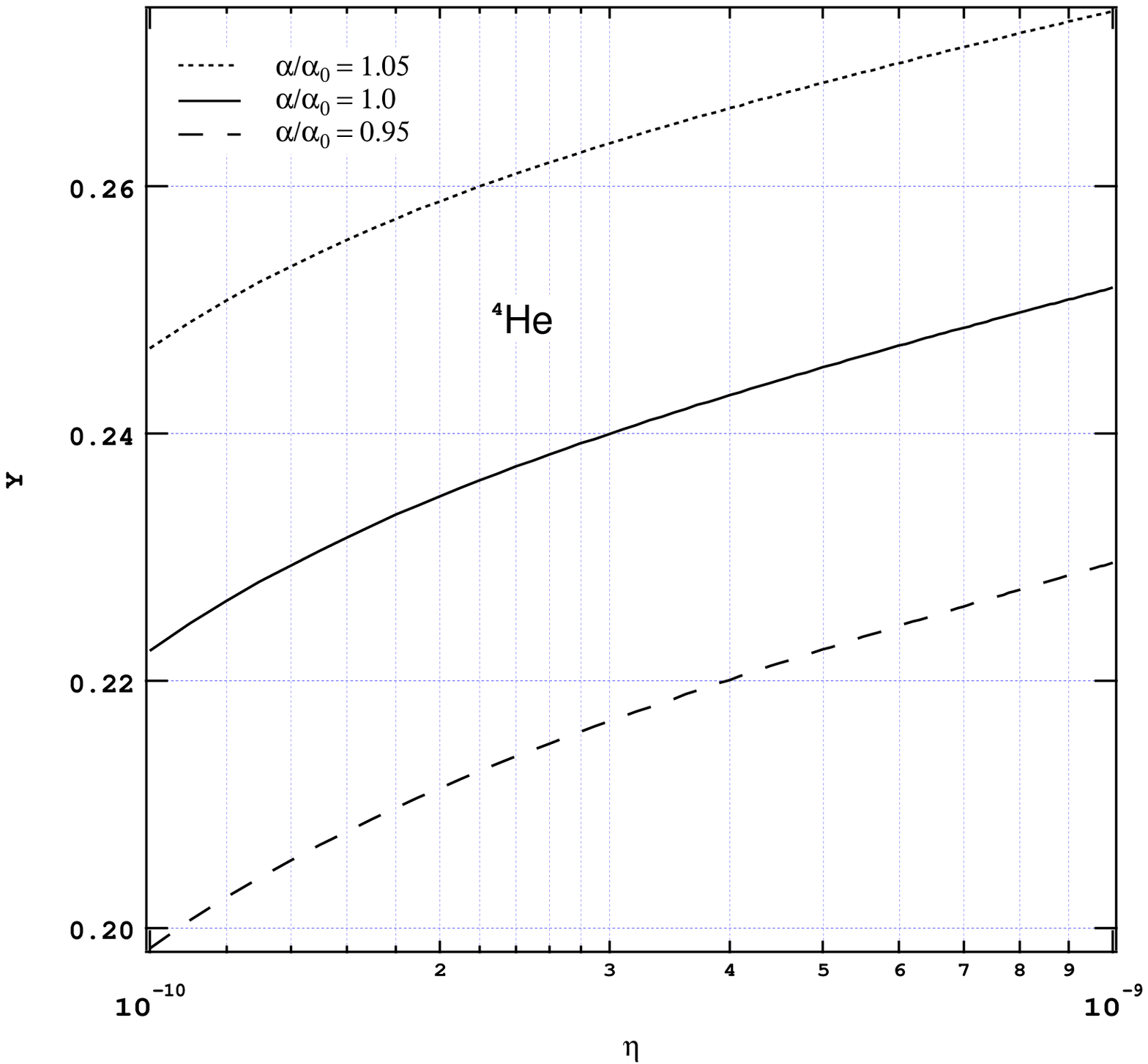,width=0.5\textwidth}\epsfig{file=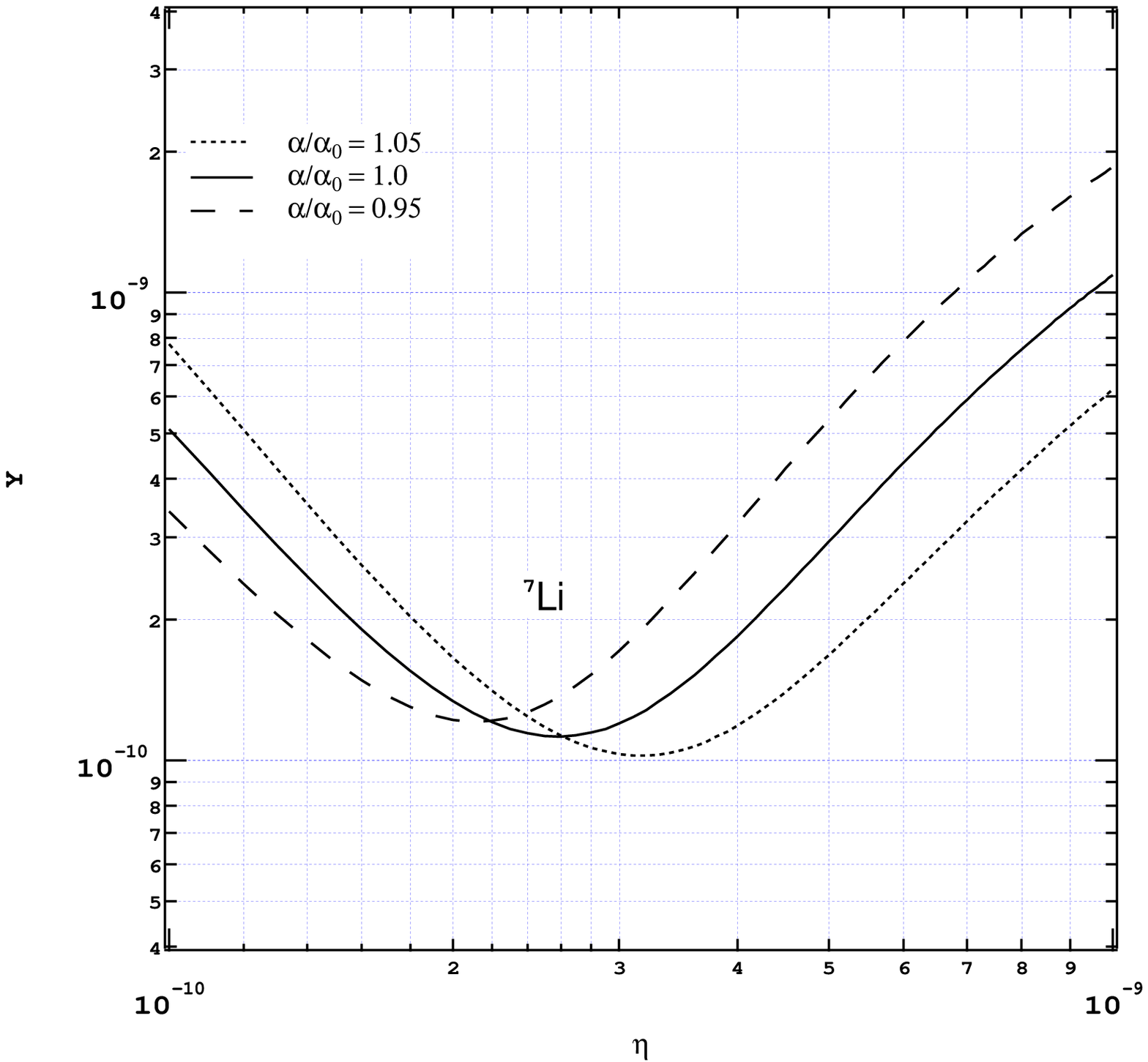,width=0.5\textwidth}}
\caption{The abundance of (a) $^4$He and (b) $^7$Li
 as a function of $\eta$, the baryon-to-photon ratio. Curves
are shown for the standard value of the fine structure constant $\alpha$,
and for a variation of $\pm$ 5 \% of the standard value.
}
\label{fig:li7vseta}
\end{figure}

In Fig.~\ref{fig:yvseta} we show the results for the final
abundances of the observationally interesting elements $^4$He,
$^3$He, $d$ and $^7$Li as a function of $\eta$, for SBBN and for a
5 \% increase or decrease of the value of $\alpha$ during
nucleosynthesis. In Figs.~\ref{fig:dvseta}--\ref{fig:li7vseta}
 the results for the abundances
are shown separately on an expanded scale. As can be seen, the
fractional change in the $^4$He abundance is quite insensitive to
the value of $\eta$, as are the other abundances with the notable
exception of $^7$Li. This is due to the two competing mechanisms
for $^7$Li production, i.e., for $\eta \lsim 3 \times 10^{-10}$, $^7$Li is
produced by $^4$He$(t,\gamma)^7$Li, a process in which the Coulomb-barrier
is not as significant (but where the change in $\Delta m$ caused by
a change in $\alpha$ is) as in the reaction $^4$He$(^3$He$,\gamma)^7$Be,
which, followed by decay of $^7$Be to $^7$Li, is the dominant reaction
that synthesizes $^7$Li for $\eta \gsim 3 \times 10^{-10}$.

\begin{figure}
\centerline{\epsfig{file=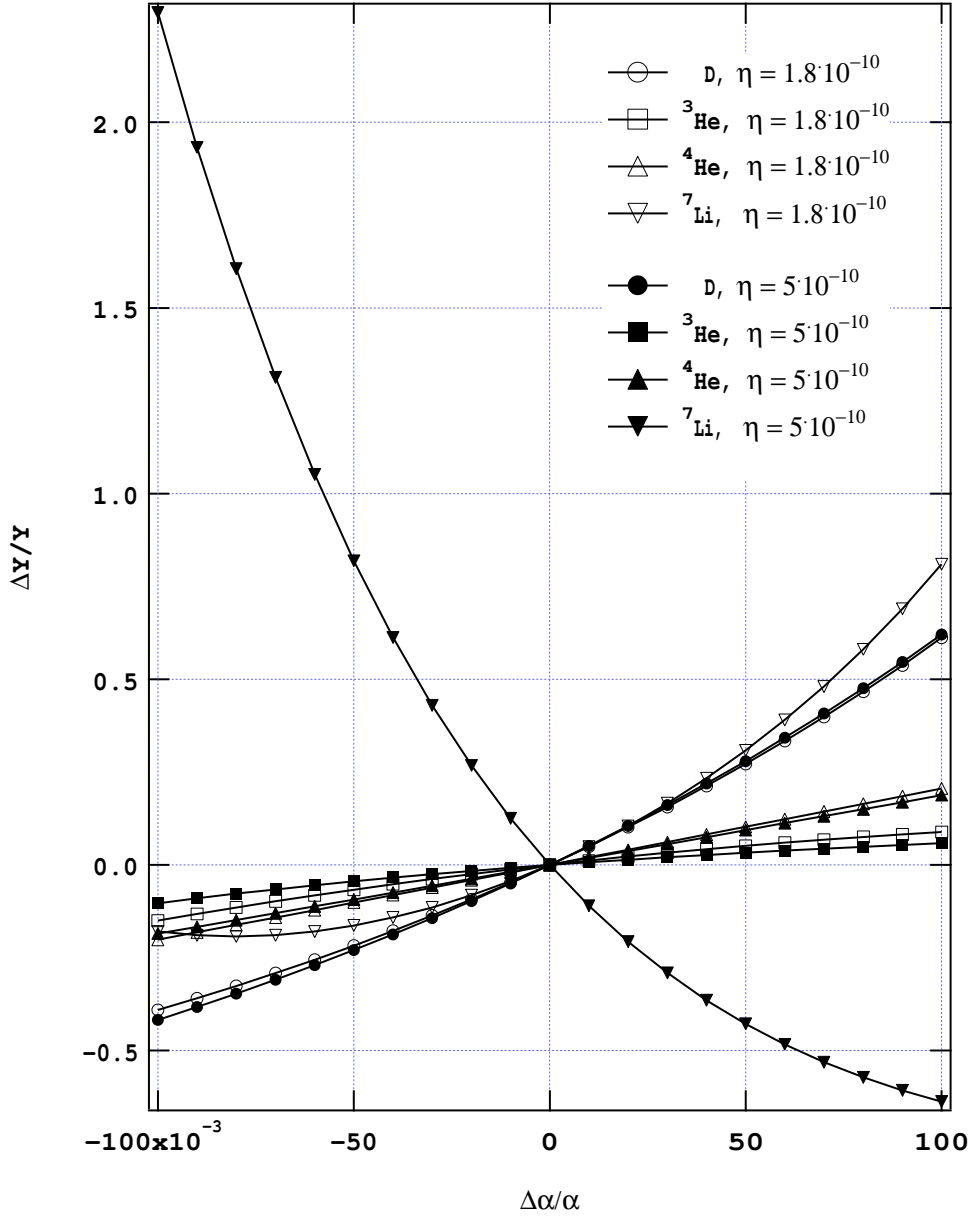,width=0.99\textwidth}}
\caption{Fractional variation of the light-element abundances $Y$ {\em vs}
fractional variation of the fine structure constant $\alpha$, for two
values of the baryon-to-photon ratio $\eta$. Notice that the curves
for $^4$He for the two $\eta$ values nearly overlap. The dramatic decrease
of the $^7$Li abundance for a high $\eta$ value is due to the strong 
Coulomb barrier in its production.}
\label{fig:dyvsda}
\end{figure}

Actually, the  of $\eta$ coming from a comparison between the predictions
of standard
nucleosynthesis and observational data 
is presently quite uncertain, as is exemplified by
the differing recent analyses of the problem in SBBN
\cite{olive,turner2}. In~\cite{olive}, a combined analysis gives
$\eta\sim 1.8\cdot 10^{-10}$ (driven largely by accepting a high
value of the deuterium abundance), whereas in~\cite{turner2} a low
deuterium abundance is taken as the preferred observational result
leading to a value $\eta\sim 5\cdot 10^{-10}$. In
Fig.~\ref{fig:dyvsda} the fractional changes in the light-element
abundances are shown versus the assumed fractional change in
$\alpha$ for both these values of $\eta$.

In the case of the low value of $\eta\sim1.8\cdot 10^{-10}$, 
the $^7$Li abundance does
not vary as strongly with a changing $\alpha$, and since this
abundance is known less accurately than that of $^4$He it is the
latter which has to be used to bound possible variations of
$\alpha$. However, the observational status $^4$He is still a
matter of debate, and it does not seem likely that the systematic
errors are larger than usually quoted. For instance, the global
fit referred to in~\cite{olive} is
\begin{equation}
\label{eq:meas4He1}
Y_4=0.238\pm 0.002\pm 0.005
\end{equation}
whereas the corresponding value in a recent reanalysis by Izotov
et al.~\cite{izotov} is given as
\begin{equation}
\label{eq:meas4He2}
Y_4=0.244\pm 0.002
\end{equation}

Using the second upper limit and the first lower limit for
the $^4$He abundance, as is difficult to avoid before better data
become available, we see from Fig.~\ref{fig:li7vseta} (a) that a
reduction of the value of $\alpha$ larger than 5 \% is not
excluded, if $^4$He is considered alone, and the other abundances are
only used to generously bound $\eta$ to be in the interval
$10^{-10}-10^{-9}$. 
Interestingly, one may note that on the contrary an increase of the
electromagnetic strength by $2$ \%
would relieve the pressure these low-$\eta$ models feel
from the upward revision of the $^4$He abundance of~\cite{izotov}, while
still falling in the allowed range for $^7$Li discussed below.
Of course, we have to remember the fact that the change in $^4$He
is caused by the model-dependent variation of the neutron-proton mass
difference with $\alpha$.

For the high value of $\eta\sim 5\cdot 10^{-10}$, we see that a
decrease of $\alpha$ by 2 \% is within our
allowed range of the $^4$He abundance (the change in deuterium caused
by such a change in $\alpha$ does not affect
very much the $\eta$ value assigned to a given deuterium measurement). 
Since for the high $\eta$
solution the variation of $^7$Li with $\alpha$ is rapid, we have
to consider the observational data for lithium. Unfortunately, the
situation is less than clear here as well. It seems that the data
are well described by a ``plateau'' value versus metallicity, given
by~\cite{bonifacio}
\begin{equation}
\label{eq:measlithium}
Y_7 = \left(1.73 \pm 0.30\right)\cdot 10^{-10}\label{eq:li}
\end{equation}

The fact that there is little dispersion around this value could
indicate that it represents the primordial abundance. However,
lithium could be destroyed by stellar processes by an unknown
amount, which has led some workers in the field~\cite{turner2} to
consider the possibility that the true primordial value is up to a
factor of 2 larger.

It is interesting to note from Fig.~\ref{fig:yvseta} that the
$^7$Li abundance for this value of $\eta$ can be brought down
to $2\cdot 10^{-10}$
if the strength of $\alpha$ was larger
by around 3 \% than the standard value (thus making the Coulomb
suppression stronger). For the same change of $\alpha$, the $^4$He
abundance is, however, violating our
(model-dependent) upper bound by a considerable amount. 
Thus, we conclude that the high-$\eta$ solution, which already has problems
explaining $^7$Li (if (\ref{eq:li}) represents the primordial value)
can not be improved by changing $\alpha$, in contrast to the 
low-$\eta$ solution which may benefit from an increase of $\alpha$ 
by a couple of percent during nucleosynthesis.

\section{Conclusions}

The successful first-order comparison between Big Bang nucleosynthesis
predictions and observed abundances implies that the values of the
fundamental coupling constants at that epoch
can not have been too much different
from the present ones. When it comes to more precise statements, the
situation is complicated by the presently partially conflicting measurements.
(Similar conclusions have recently been obtained concerning possible
limits on the effective number of neutrino species during nucleosynthesis
\cite{lisi}.)
We argue that it is impossible to quote a better bound than $2$ \% on
the deviation from the present value of $\alpha$ at the time of 
nucleosynthesis, and note that such a variation may even be implied,
if the deuterium observations leading to the low-$\eta$ solution are confirmed
with increased significance, and the trend continues of an increasing 
value for $^4$He from observations.

A new ingredient in our treatment is the discovery of the large sensitivity,
especially at high values of $\eta$, of the $^7$Li abundance on
the electromagnetic strength, due to exponential Coulomb
barriers. If the experimental and theoretical situation is improved for
this isotope, it holds the promise of giving the most stringent bound
on variation of $\alpha$, with the added virtue that it is less sensitive
to model assumptions than, e.g., $^4$He. The problem that the
high-$\eta$ (low deuterium) solution may have in explaining the low
plateau value of $^7$Li can in principle be solved by an increase in
$\alpha$
by a few percent, but only if the true primordial $^4$He abundance
is well above $0.25$ (or if $\Delta m$ depends on $\alpha$ in a different
way than we have assumed).

It is of interest to notice that other exotic effects such as the 
presence of strong primordial magnetic
fields mainly affect the $^4$He abundance~\cite{Grasso}  and not Li or
the other elements.
Also in this respect the signal given by $^7$Li is clearer.

The limit obtained here is about the same as the one potentially
given by the recombination time test \cite{cmbr}, but probably more
significant since it is at a higher redshift. 
However, at the time when the microwave background was emitted, the
universe was  totally dominated by gravity and electromagnetic
interactions from which it can be argued  that it offers a less
model-dependent test.

\section{Acknowledgements}
We thank Roberto Liotta for interesting discussions on the nuclear physics
aspects of this problem. L.B. thanks the Swedish Natural Science
Research Council for support, and S.I. is grateful to the Swedish Institute
for a scholarship.


\pagebreak

\begin{table}
\scriptsize
\begin{tabular}{|c|c|l|}
\hline
\textbf{Number} & \textbf{Reaction} & {\hspace{4.5cm} \textbf{Reaction
rate}} \\
\hline
\hline
$1$ & $^2H(p,\gamma)^3He$ & $2.650 \times 10^3 (1+\delta)^{\frac{1}{3}}
T_9^{-\frac{2}{3}} exp(-3.720 (1+\delta)^{\frac{2}{3}} T_9^{-\frac{1}{3}})
\times$ \\
  &   & $\times (1.000 + 1.120 \times 10^{-1} (1+\delta)^{-\frac{2}{3}}
T_9^{\frac{1}{3}} + 1.990 (1+\delta)^{\frac{2}{3}} T_9^{\frac{2}{3}} +
1.560
T_9 +$ \\
  &   & $+ 1.620 \times 10^{-1} (1+\delta)^{\frac{4}{3}} T_9^{\frac{4}{3}}
+
3.240 \times 10^{-1} (1+\delta)^{\frac{2}{3}} T_9^{\frac{5}{3}} )$ \\
\hline
$2$ & $^3H(p,\gamma)^4He$ & $2.200 \times 10^4 (1+\delta)^{\frac{1}{3}}
T_9^{-\frac{2}{3}} exp(-3.869 (1+\delta)^{\frac{2}{3}} T_9^{-\frac{1}{3}})
\times$ \\
  &   & $\times ( 1.000 + 1.080 \times 10^{-1} (1+\delta)^{-\frac{2}{3}}
T_9^{\frac{1}{3}} + 1.680 (1+\delta)^{\frac{2}{3}} T_9^{\frac{2}{3}} +
1.260
T_9 +$ \\
  &   & $+ 5.510 \times 10^{-1} (1+\delta)^{\frac{4}{3}} T_9^{\frac{4}{3}}
+
1.060 (1+\delta)^{\frac{2}{3}} T_9^{\frac{5}{3}} )$ \\
\hline
$3$ & $^6Li(p,\gamma)^7Be$ & $6.690 \times 10^5 (1+\delta)^{\frac{1}{3}}
T_{9a}^{\frac{5}{6}} T_9^{-\frac{3}{2}} exp(-8.413
(1+\delta)^{\frac{2}{3}}
T_{9a}^{-\frac{1}{3}})$ \\
  &   & $\left[ T_{9a}=T_9/(1.000-9.690 \times 10^{-2} T_9 + 2.000
(1+\delta)^{\frac{2}{3}}T_9^{\frac{5}{3}}(1.000-9.690 \times 10^{-2}
T_9)^{-\frac{2}{3}}) \right]$ \\
\hline
$4$ & $^6Li(p,\alpha)^3He$ & $3.730 \times 10^{10}
(1+\delta)^{\frac{1}{3}}
T_9^{-\frac{2}{3}} exp(-8.413 (1+\delta)^{\frac{2}{3}} T_9^{-\frac{1}{3}}
-
(1.818 \times 10^{-1} T_9)^2 (1+\delta)^2) \times$ \\
  &   & $\times ( 1.000 + 5.000 \times 10^{-2} (1+\delta)^{-\frac{2}{3}}
T_9^{\frac{1}{3}} - 6.100 \times 10^{-2} (1+\delta)^{\frac{2}{3}}
T_9^{\frac{2}{3}} -$\\
  &   & $- 2.100 \times 10^{-2} T_9 + 6.000 \times 10^{-3}
(1+\delta)^{\frac{4}{3}} T_9^{\frac{4}{3}} +
5.000 \times 10^{-3} (1+\delta)^{\frac{2}{3}} T_9^{\frac{5}{3}} ) +$ \\
  &   & $+ 1.330 \times 10^{10} T_9^{-\frac{3}{2}} exp(-1.776 \times
10^{1} T_9^{-1}) +$ \\
  &   & $+1.290 \times 10^{9} T_9^{-1} exp(-2.182 \times 10^{1} T_9^{-1})$
\\
\hline
$5$ & $^7Li(p,\alpha)^4He$ & $1.096 \times 10^9 (1 +
\delta)^{\frac{1}{3}}T_9^{-\frac{2}{3}}
exp(-8.472 (1+\delta)^{\frac{2}{3}} T_9^{-\frac{1}{3}}) -$ \\
  &   & $- 4.830 \times 10^8 (1+\delta)^{\frac{1}{3}} T_{9b}^{\frac{5}{6}}
T_9^{-\frac{3}{2}} exp(-8.472 (1+\delta)^{\frac{2}{3}}
T_{9b}^{-\frac{1}{3}})
+$ \\
  &   & $+ 1.060 \times 10^{10} T_9^{-\frac{3}{2}} exp(-3.044 \times
10^{1} T_9^{-1}) +$ \\
  &   & $+ 1.560 \times 10^5 (1+\delta)^{\frac{1}{3}} T_9^{-\frac{2}{3}}
exp((-8.472 (1+\delta)^{\frac{2}{3}} T_9^{-\frac{1}{3}}) - (5.896 \times
10^{-1} T_9)^2  (1+\delta)^2) \times$ \\
  &   & $\times ( 1.000 + 4.900 \times 10^{-2} (1+\delta)^{-\frac{2}{3}}
T_9^{\frac{1}{3}} - 2.498 (1+\delta)^{\frac{2}{3}} T_9^{\frac{2}{3}} +$ \\
  &   & $+ 8.600 \times 10^{-1} T_9 + 3.518 (1+\delta)^{\frac{4}{3}}
T_9^{\frac{4}{3}} + 3.080 (1+\delta)^{\frac{2}{3}} T_9^{\frac{5}{3}} ) +$
\\
  &   & $+ 1.550 \times 10^{6} T_9^{-\frac{3}{2}} exp(-4.478
T_9^{-1})$ \\
  &   & $\left[ T_{9b}   = T_9/(1.000+0.759T_9) \right]$\\
\hline
$6$ & $^2H(\alpha,\gamma)^6Li$ & $3.010 \times 10^{1}
(1+\delta)^{\frac{1}{3}}
T_9^{-\frac{2}{3}} exp(-7.423 (1+\delta)^{\frac{2}{3}} T_9^{-\frac{1}{3}})
\times$ \\
  &   & $\times ( 1.000 + 5.600 \times 10^{-2} (1+\delta)^{-\frac{2}{3}}
T_9^{\frac{1}{3}} - 4.850 (1+\delta)^{\frac{2}{3}} T_9^{\frac{2}{3}} +
8.850
T_9 -$ \\
  &   & $- 5.850 \times 10^{-1} (1+\delta)^{\frac{4}{3}} T_9^{\frac{4}{3}}
-
5.840 \times 10^{-1} (1+\delta)^{\frac{2}{3}} T_9^{\frac{5}{3}} )+$ \\
  &   & $+ 8.550 \times 10^1 T_9^{-\frac{3}{2}} exp(-8.228
T_9^{-1})$ \\
\hline
\end{tabular}
\normalsize
\caption{{Reaction rates (in units of cm$^3$s$^{-1}$mole$^{-1}$)
for the most relevant $\alpha$-dependent
reactions involved in SBBN (reactions 1-6). $T_9$ is the temperature
in units of $10^9$ Kelvin.}
 \label{tab:summary}
}
\end{table}
\newpage
\begin{table}
\scriptsize
\begin{tabular}{|c|c|l|}
\hline
\bf{Number} & \bf{Reaction} & {\hspace{4.2cm} \bf{Reaction rate}} \\
\hline
\hline
$7$ & $^3H(\alpha,\gamma)^7Li$ & $3.032 \times 10^5
(1+\delta)^{\frac{1}{3}}
T_9^{-\frac{2}{3}} exp(-8.090 (1+\delta)^{\frac{2}{3}} T_9^{-\frac{1}{3}})
\times$ \\
  &   & $\times ( 1.000 + 5.160 \times 10^{-2} (1+\delta)^{-\frac{2}{3}}
T_9^{\frac{1}{3}} + 2.290 \times 10^{-2} (1+\delta)^{\frac{2}{3}}
T_9^{\frac{2}{3}} +$ \\
  &   & $ + 8.280 \times 10^{-3} T_9 - 3.280 \times 10^{-4}
(1+\delta)^{\frac{4}{3}} T_9^{\frac{4}{3}} - 3.010 \times 10^{-4}
(1+\delta)^{\frac{2}{3}} T_9^{\frac{5}{3}} ) +$ \\
  &   & $+ 5.109 \times 10^5 (1+\delta)^{\frac{1}{3}} T_{9c}^{\frac{5}{6}}
T_9^{-\frac{3}{2}} exp(-8.068 (1+\delta)^{\frac{2}{3}}
T_{9c}^{-\frac{1}{3}})$
\\
  &   & $\left[ T_{9c}   = T_9/(1.000+0.138T_9) \right]$\\
\hline
$8$ & $^3He(\alpha,\gamma)^7Be$ & $4.817 \times 10^{6}
(1+\delta)^{\frac{1}{3}} T_9^{-\frac{2}{3}} exp(-1.496 \times 10^{1}
(1+\delta)^{\frac{2}{3}} T_9^{-\frac{1}{3}}) \times$ \\
  &   & $\times ( 1.000 + 3.250 \times 10^{-2} (1+\delta)^{-\frac{2}{3}}
T_9^{\frac{1}{3}} - 1.040 \times 10^{-3} (1+\delta)^{\frac{2}{3}}
T_9^{\frac{2}{3}} -$ \\
  &   & $- 2.370 \times 10^{-4} T_9 - 8.110 \times 10^{-5}
(1+\delta)^{\frac{4}{3}} T_9^{\frac{4}{3}} - 4.690 \times 10^{-5}
(1+\delta)^{\frac{2}{3}} T_9^{\frac{5}{3}} ) +$ \\
  &   & $+ 5.938 \times 10^{6} (1+\delta)^{\frac{1}{3}}
T_{9d}^{\frac{5}{6}}
T_9^{-\frac{3}{2}} exp(-1.286 \times 10^{1} (1+\delta)^{\frac{2}{3}}
T_{9d}^{-\frac{1}{3}})$ \\
  &   & $\left[ T_{9d}   = T_9/(1.000+0.107T_9) \right]$\\
\hline
$9$ & $^2H(d,n)^3He$ & $3.950 \times 10^8 (1+\delta)^{\frac{1}{3}}
T_9^{-\frac{2}{3}} exp(-4.259 (1+\delta)^{\frac{2}{3}} T_9^{-\frac{1}{3}})
\times$ \\
  &   & $\times (1.000 + 9.800 \times 10^{-2} (1+\delta)^{-\frac{2}{3}}
T_9^{\frac{1}{3}} + 7.650 \times 10^{-1} (1+\delta)^{\frac{2}{3}}
T_9^{\frac{2}{3}} +$ \\
  &   & $ + 5.250 \times 10^{-1} T_9 + 9.610 \times 10^{-3}
(1+\delta)^{\frac{4}{3}} T_9^{\frac{4}{3}} + 1.670 \times 10^{-2}
(1+\delta)^{\frac{2}{3}} T_9^{\frac{5}{3}} )$ \\
\hline
$10$ & $^2H(d,p)^3H$ & $4.170 \times 10^8 (1+\delta)^{\frac{1}{3}}
T_9^{-\frac{2}{3}} exp(-4.258 (1+\delta)^{\frac{2}{3}} T_9^{-\frac{1}{3}})
\times$ \\
  &   & $\times ( 1.000 + 9.800 \times 10^{-2} (1+\delta)^{-\frac{2}{3}}
T_9^{\frac{1}{3}} + 5.180 \times 10^{-1} (1+\delta)^{\frac{2}{3}}
T_9^{\frac{2}{3}} +$ \\
  &   & $ + 3.550 \times 10^{-1} T_9 - 1.000 \times 10^{-2}
(1+\delta)^{\frac{4}{3}} T_9^{\frac{4}{3}} - 1.800 \times 10^{-2}
(1+\delta)^{\frac{2}{3}} T_9^{\frac{5}{3}} )$ \\
\hline
$11$ & $^3H(d,n)^4He$ & $1.063 \times 10^{11} (1+\delta)^{\frac{1}{3}}
T_9^{-\frac{2}{3}} exp(-4.559 (1+\delta)^{\frac{2}{3}} T_9^{-\frac{1}{3}}
-
(1.326 \times 10^{1} T_9)^2 (1+\delta)^2) \times$ \\
  &   & $\times ( 1.000 + 9.200 \times 10^{-2} (1+\delta)^{-\frac{2}{3}}
T_9^{\frac{1}{3}} - 3.750 \times 10^{-1} (1+\delta)^{\frac{2}{3}}
T_9^{\frac{2}{3}} -$ \\
  &   & $ - 2.420 \times 10^{-1} T_9 + 3.382 \times 10^{1}
(1+\delta)^{\frac{4}{3}} T_9^{\frac{4}{3}} + 5.542 \times 10^{1}
(1+\delta)^{\frac{2}{3}} T_9^{\frac{5}{3}} ) +$ \\
  &   & $+ 8.047 \times 10^8 T_9^{-\frac{2}{3}} exp(-0.486
T_9^{-1})$ \\
\hline
$12$ & $^3He(d,p)^4He$ & $5.021 \times 10^{10} (1+\delta)^{\frac{1}{3}}
T_9^{-\frac{2}{3}} exp(-7.144 (1+\delta)^{\frac{2}{3}} T_9^{-\frac{1}{3}}
-
(3.704 T_9)^2 (1+\delta)^2) \times$ \\
  &   & $\times ( 1.000 + 5.800 \times 10^{-2} (1+\delta)^{-\frac{2}{3}}
T_9^{\frac{1}{3}} + 6.030 \times 10^{-1} (1+\delta)^{\frac{2}{3}}
T_9^{\frac{2}{3}} +$ \\
  &   & $ + 2.450 \times 10^{-1} T_9 + 6.970 (1+\delta)^{\frac{4}{3}}
T_9^{\frac{4}{3}} + 7.190 (1+\delta)^{\frac{2}{3}} T_9^{\frac{5}{3}} ) +$
\\
  &   & $+ 5.212 \times 10^8 T_9^{-\frac{1}{2}} exp(-1.762
T_9^{-1})$ \\
\hline
\end{tabular}
\normalsize
\caption{{Reaction rates (in units of cm$^3$s$^{-1}$mole$^{-1}$)
for the most relevant $\alpha$-dependent
reactions involved in SBBN(reactions 7-12).  $T_9$ is the temperature
in units of $10^9$ Kelvin.} \label{tab:summary2}}
\end{table}


\end{document}